\documentclass[12pt]{article}
{\begin{Sbox}\begin{minipage}}%
		{\end{minipage}\end{Sbox}\fbox{\TheSbox}}%
\usepackage[]{amssymb}
\usepackage{amsmath, amsthm, anysize, enumerate, color, url}
\allowdisplaybreaks[4]
\usepackage{graphicx}
\usepackage{epstopdf}
\usepackage{epsfig}
\usepackage{subfigure}
\usepackage{multirow}
\usepackage{tikz}
\usepackage{array}

\usepackage{algorithm}
\usepackage{algpseudocode}
\usepackage[sectionbib]{natbib}

\usepackage{threeparttable}

\usepackage[colorlinks, breaklinks,
linkcolor=red,
citecolor=blue
]{hyperref}
\usepackage{bm}

\newtheorem{theorem}{Theorem} %%%[section]
\newtheorem{corollary}{Corollary}

\newtheorem{lemma}{Lemma} %%%[theorem]
\newcommand{\balign}{\begin{align}}
	\newcommand{\ealign}{\end{align}}
\newcommand{\beq}{\begin{equation}}
	\newcommand{\eeq}{\end{equation}}

\theoremstyle{definition}
\newtheorem{remark}{Remark}  %%%[theorem]

\renewcommand{\P}{\mathbb{P}}

\newcommand{\goto}{\rightarrow}

\usepackage{setspace}
\allowdisplaybreaks
\usepackage{xr}
\begin{document}
\title{\bf Triple-dyad ratio estimation for the $p_1$ model\footnote{All authors contribute equally.}}

\author{Qunqiang Feng\textsuperscript{1}
	\hspace{4mm} \and
	Yaru Tian\textsuperscript{1}
	\hspace{4mm} \and
	Ting Yan\textsuperscript{2}
}

\footnotetext[1]{Department of Statistics, University of Science and Technology of China, Hefei, 230000, China.
	\texttt{Email:} fengqq@ustc.edu.cn, tt1997@mail.ustc.edu.cn}
\footnotetext[2]{Department of Statistics, Central China Normal University, Wuhan, 430079, China.
	\texttt{Email:} tingyanty@mail.ccnu.edu.cn}
\date{}

\maketitle

\begin{abstract}
	\begin{spacing}{1.2}
	Although the $p_1$ model was proposed 40 years ago, little progress has been made to address asymptotic theories in this model, that is, neither consistency of the maximum likelihood estimator (MLE) nor other parameter estimation with statistical guarantees is understood. This problem has been acknowledged as a long-standing open problem. To address it, we propose a novel parametric estimation method based on the ratios of the sum of a sequence of triple-dyad indicators to another one, where a triple-dyad indicator means the product of three dyad indicators.
	Our proposed estimators, called \emph{triple-dyad ratio estimator}, have explicit expressions and can be scaled to very large networks with millions of nodes.
	We establish the consistency and asymptotic normality of the triple-dyad ratio estimator when the number of nodes reaches infinity.
	Based on the asymptotic results, we develop a test statistic for evaluating whether is a reciprocity effect in directed networks. 
	The estimators for the density and reciprocity parameters contain bias terms, where analytical bias correction formulas are proposed to make valid inference.
	Numerical studies demonstrate the findings of our theories and show that the estimator is comparable to the MLE in large networks.
	\end{spacing}

	\begin{spacing}{1.4}
	\end{spacing}
	
\end{abstract}

\section{Introduction}

The $p_1$ model, proposed by \cite{Holland1981exponential}, is an early statistical model for modelling network data. It is an exponential family of probability distribution for directed graphs with in- and out-degrees and the number of reciprocated edges as sufficient statistics. The $p_1$ model assumes that the probability of the occurrence of a directed graph $G_n$ with $n$ nodes is
\begin{align*}
	\P(G_n = g_n ) \propto
	\exp\Bigg\{\rho m+\theta x_{++}+\sum\limits_{i=1}^n\alpha_ix_{i+}+\sum\limits_{j=1}^n\beta_{j}x_{+j}\Bigg\},
\end{align*}
where $m$, $x_{++}$, $x_{i+}$ and $x_{+j}$ are the number of reciprocated edges, total number of edges, out-degree of node $i$ and in-degree of node $j$ in the observed graph $g_n$, respectively. The parameters $\rho$, $\theta$, $\{\alpha_i\}_{i=1}^n$, and $\{\beta_j\}_{j=1}^n$ quantify the force of reciprocated or mutual edges between pairs of nodes, density of graphs, and expansiveness and popularity across different nodes, respectively. In brief, the $p_1$ model is a simple but powerful tool for characterizing tendencies toward reciprocation and differential attraction.

Since its conception, the $p_1$ model has been widely applied in practice [e.g., \cite{Iacobucci1992,Igarashi2005,OMalley2008TheAO}] and continues to form the foundation of many other models for network analysis [e.g., \cite{Wasserman1996logit,Robins2007173}]. Despite its apparent simplicity and popularity, the $p_1$ model poses big challenges for asymptotic theory.
\cite{Fienberg2012} points out that
``\emph{Given that the basic network models described in Section 3 (including $p_1$ model) have been in use for several
	decades, it is surprising that their statistical properties did not develop somehow in parallel.}
"
Furthermore,
\cite{Goldenberg2010} states that

``\emph{A major problem with the $p_1$ and related models, recognized by Holland and Leinhardt, is the lack of standard asymptotics to assist in the development of goodness-of-fit procedures for the model. Since the number of $\{\alpha_i\}$ and $\{\beta_j\}$ increases directly with the number of nodes,
	we have no consistency in results for the maximum likelihood estimates, and no simple way to test for $\rho=0$.}''

Since large sample theory cannot be applied to the $p_1$ and related models, as \cite{Fienberg1981}, \cite{Goldenberg2010} and \cite{Fienberg2012} argue, statisticians have long explored asymptotic theories in these models.
One possibility is a simpler $\beta$-model for undirected graphs, a name coined by \cite{Chatterjee2011random}; this model may be considered an undirected version of the $p_1$ model. \cite{Chatterjee2011random} established uniform consistency of the maximum likelihood estimator (MLE) in the $\beta$-model, while \cite{Yan2013clt} derived its central limit theorem. \cite{Rinaldo2013} derived the conditions of MLE existence in both the $\beta$-model and $p_1$ model. \cite{Yan2016exponential} further established uniform consistency and asymptotic normalities of the MLE in
the $p_0$ model, a name coined by \cite{Yan2021sinica} later, which was a special case of the $p_1$ model without the reciprocity parameter.
Statisticians have also derived asymptotic theories in some  generalized $\beta$-models \cite[e.g.][]{Perry:Wolfe:2012,Hillar:Wibisono:2013,Graham2017,Yan-Jiang2019,Chen:2020,chang2024,shao2023l2}. \cite{Yan2015p1}, for instance, conducted simulation studies of MLE performance in the $p_1$ model. \cite{Fienberg2011p1} and \cite{Petrovic2010algebraic} studied its geometric properties from the perspective of algebraic statistics.

Contrary to the $p_0$ model studied by \cite{Yan2016exponential},
another reciprocity parameter exists that implies non-independence of edges. This makes asymptotic inference in the $p_1$ model more challenging.
It is remarkable that \cite{Chatterjee2011random} obtained the convergence rate of a fixed-point iterative algorithm for proving consistency,
and \cite{Yan2013clt} and \cite{Yan2016exponential} constructed a simple matrix to approximate the inverse of the Fisher information matrix
for deriving asymptotic properties of the MLE.
These techniques depend crucially on either the diagonally balanced or dominant properties of the  Fisher information matrix,
which are implied by the independence of edges.
Such properties do not hold in the $p_1$ model due to the appearance of reciprocity parameter (\cite{Yan2015p1}).
Therefore, it is unclear how to adapt their approaches to the $p_1$ model.
This has slowed the progress of the asymptotic theory of the $p_1$ model, in that neither consistency of the MLE nor other parameter estimation with statistical guarantees is understood, giving us a long-standing open problem [e.g., \cite{Goldenberg2010,Fienberg2012}].

To address our open problem, we propose a novel parametric estimation method based on the ratios of the sum of a sequence of triple-dyad indictors to another one,
where a triple-dyad indicator means the product of triple indicators ($I_{ij}^{a_1,b_1}, I_{jk}^{a_2,b_2}, I_{ki}^{a_3,b_3}$), and $I_{ij}^{ab}=(X_{ij}=a, X_{ji}=b)$ indicates the binary edge values between nodes $i$ and $j$.
Here, $X_{ij}\in\{0,1\}$ denotes whether there exists one directed edge from $i$ to $j$.
The estimators of the parameters $\theta$, $\rho$, $\alpha_i$, and $\beta_j$ are provided in next section (see \eqref{thetahat}, \eqref{rhohat}, \eqref{alphaihat} and \eqref{betajhat}, respectively). Our central idea is that these unknown parameters or their linear combinations can be represented as the form of the logarithm of ratios of $p_{ij}^{a_1,b_1} p_{jk}^{a_2,b_2} p_{ki}^{a_3,b_3}$ to $p_{ij}^{a_4,b_4}p_{jk}^{a_5,b_5} p_{ki}^{a_6,b_6}$, where $p_{ij}^{a_k,b_k}$ denotes the probability of observing the configuration of a dyad $(X_{ij}=a_k, X_{ji}=b_k)$ and $a_k,b_k \in\{0,1\}$. We can then use empirical values $I_{ij}^{ab}$ to estimate these unknown dyadic probabilities.

Our proposed estimators, called \emph{triple-dyad ratio estimator}, have explicit expressions, unlike the MLE, which has to resort to iterative algorithms because its explicit solution is not possible. Therefore, our estimation can be scaled to very large networks with millions of nodes as long as the memory of the computer is large enough, while the maximum likelihood estimation becomes useless in many network models owing to computational challenge.
We establish the consistency and asymptotic normality of the triple-dyad ratio estimator when the number of nodes reaches infinity.
Their proofs need to address complex dependence amongst triple-dyad indicators.
To resolve this issue, we develop decomposing techniques that, by introducing intermediate variables, divide the sum of triple-dyad indicators into several parts consisting of independent or conditional independent random variables.
The calculations of asymptotic variances of estimators are highly nontrivial since it is involved with the analysis of hundreds of terms.
In addition, the estimators for the density and reciprocity parameters are not unbiased,
which is collaborated with the finding in the simulation studies of \cite{Yan2015p1} for evaluating asymptotic distribution of the MLE.
We provide analytical bias correction formulas to make valid inference.
We conduct numerical studies to demonstrate the findings of our theory and show that the estimator is comparable to the MLE in large networks.	

The rest of the paper is organized as follows. Section \ref{section-estimation} introduces the estimation method. Section \ref{section-theory} presents the consistency and asymptotic normality of the triple-dyad ratio estimator. Section \ref{section-test} presents the hypothesis testing methods for unknown parameters.
Section \ref{section-simulation} presents the simulation studies. Section \ref{section-discussion} discusses our findings and concludes the paper.
All the technical proofs are relegated into Supplementary Material.

We close this section by introducing some notation that will be used in the paper. For a positive integer $n$, $[n]$ denotes the set $\{1, \ldots, n\}$ and ${\bm I}_{n}$ is an identity matrix of order $n$. We use $c_1, c_2, \ldots $ to denote absolute constants with different values in different places. For a random event $E$, let $I(E)$ be its indicator. For a vector ${\bm x}=(x_1,\ldots,x_n)^{\top}\in\mathbb{R}^n$, denote by $\|{\bm x}\|_{\infty}=\max_{1\le i\le n}|x_i|$ the $l_{\infty}$-norm of ${\bm x}$, and ${\rm diag}({\bm x})$ is a square diagonal matrix with ${\bm x}$ as the main diagonal. For any sequences $\{a_n\}_{n\geq 1}$ and $\{b_n\}_{n\geq 1}$,
we say that $a_n\lesssim b_n$ or $a_n=O( b_n)$ if $a_n \le c b_n$ for some absolute constant $c$. 
We write $a_n\ll b_n$ or $a_n=o( b_n)$ if $a_n/b_n \to 0$. If $a_n\lesssim b_n$ and $a_n\gtrsim b_n$, we write $a_n\asymp b_n$.

\section{Method of triple-dyad ratio estimation}
\label{section-estimation}

Recall that the graph $G_n$ contains $n$ nodes labelled ``$1,2, \ldots, n$'' and ${\bm X} =(X_{ij})_{i,j=1}^n$ denotes the adjacency matrix of $G_n$, where $X_{ij}$ denotes whether there exists a directed edge from node $i$ to node $j$ and $X_{ii}=0$ by convention. That is, $X_{ij}=1$ indicates a directed edge from tail node $i$ to head node $j$; otherwise $X_{ij}=0$.
For each pair $(i,j)$, there exist four possible dyadic configurations for $(X_{ij}, X_{ji})$: null, $(0, 0)$; mutual or reciprocal, $(1, 1)$; or asymmetric, $(0, 1)$ and $(1, 0)$. Denote by $D_{ij}=(X_{ij},X_{ji})$ the dyad of edge indicator variables $X_{ij}$ and $X_{ji}$ between two distinct nodes $i$ and $j$.

The $p_1$ model implies that all $n(n-1)/2$ dyads are mutually independent and the probability distribution of each dyad $D_{ij}~(1\le i<j\le n)$ is

\begin{align}\label{pijab}	p_{ij}^{ab}:= \mathbb{P}(D_{ij}=(a,b))=\frac{1}{k_{ij}}\exp\big\{a(\theta+\alpha_{i}+\beta_{j})+b(\theta+\alpha_{j}+\beta_{i})+ab\rho\big\},
\end{align}
where $a,b\in\{0,1\}$ and the normalizing constant $k_{ij}$ is
\begin{equation}\label{def-kij}
	k_{ij}=1+\exp \{\theta+\alpha_{i}+\beta_{j}\}+\exp \{\theta+\alpha_{j}+\beta_{i}\}+\exp \{\rho+2 \theta+\alpha_{i}+\alpha_{j}+\beta_{i}+\beta_{j}\}.
\end{equation}
As mentioned before, 
the parameter $\rho$ measures the average tendency toward reciprocation for all pairs of nodes, while $\theta$ is a density parameter governing the sparsity of the directed graph $G_n$. $\alpha_i$ is an intrinsic expansiveness parameter quantifying the effect of an outgoing edge from node $i$. If $\alpha_i$ is large and positive, node $i$ will appear to produce more outgoing edges such that it tends to have a relatively large out-degree. Contrariwise, $\beta_j$ is a popularity parameter quantifying the effect of an incoming edge connecting to node $j$. If $\beta_j$ is large and positive, node $j$ will appear to attract more incoming edges from other nodes such that it has a relatively large in-degree.

Note that there exist $2n+2$ unknown parameters in the $p_1$ model. We collect them as an unknown parameter vector
\begin{align}
	\label{definition-Theta}
	\Theta=(\rho,\theta,\alpha_1,\ldots,\alpha_{n},\beta_1,\ldots,\beta_{n})^{\top}.
\end{align}
Because adding a constant to $\alpha_i$ and subtracting the same constant from $\theta$ leads to the same probability in \eqref{pijab}, $\{\alpha_i\}_{i=1}^n$ needs to be restricted for the identification of the model. We apply the same reasoning to $\{\beta_j\}_{j=1}^n$. Following \cite{Holland1981exponential}, we use the restrictions
\begin{equation}\label{constraint}
	\sum_{i=1}^n \alpha_i =0, \quad \sum_{j=1}^n \beta_j =0,
\end{equation}
as the conditions of model identification.

We now present the estimation method for the unknown parameters. It is remarkable that the sum of several parameters can be expressed as the logarithm of the ratio of the product of triple probabilities to another one. Specifically, for any three distinct nodes $i,j,t\in [n]$, we have
\begin{eqnarray}
	\nonumber
	\log \frac{p_{it}^{01}p_{ij}^{00}p_{tj}^{01}}{p_{it}^{00}p_{ij}^{01}p_{tj}^{00}}
	& = & 	\log \frac{\tilde{p}_{it}^{01}\tilde{p}_{ij}^{00}\tilde{p}_{tj}^{01} }{ \tilde{p}_{it}^{00}\tilde{p}_{ij}^{01}\tilde{p}_{tj}^{00} }
	\\
	\label{thetaabk}
	& =	& (\theta+\alpha_t + \beta_i) + (\theta+\alpha_j+\beta_t) - (\theta + \alpha_j + \beta_i) =
	\theta+\alpha_t+\beta_t,
\end{eqnarray}
where $\tilde{p}_{ij}^{ab}= k_{ij}\cdot p_{ij}^{ab}$.
Here, $p_{it}^{01}p_{ij}^{00}p_{tj}^{01}$ is the probability of observing the subgraph structure $j \to t \to i$ of triad $(i,j,t)$, and $p_{it}^{00}p_{ij}^{01}p_{tj}^{00}$ is the probability of the occurrence of the directed edge from $j$ to $i$ and no other edges amongst the triad $(i,j,t)$. This motivates an empirical estimator for the density parameter $\theta$.

For any $a,b \in \{0,1\}$ and $i,j\in [n]$, define the indicator
\begin{align*}
	I_{ij}^{ab}:=I(D_{ij}=(a,b)), \quad i\neq j,
\end{align*}
which is a Bernoulli random variable with success rate $p_{ij}^{ab}$.
According to the restriction in \eqref{constraint}, summing $t$ in \eqref{thetaabk} yields an empirical estimator for $\theta$, given by
\begin{align}\label{thetahat}
	\hat{\theta}=\frac{1}{n}\sum\limits_{t=1}^n\log\frac{\sum_{i,j\neq t}I_{it}^{01}I_{ij}^{00}I_{tj}^{01}}{\sum_{i,j\neq t}I_{it}^{00}I_{ij}^{01}I_{tj}^{00}}.
\end{align}
where the sum $\sum_{i,j\neq t}$ is taken over all distinct nodes $i,j\in[n]$ and $i,j\neq t$.
The numerator in the above logarithmic term counts the number of ordered pairs of nodes $(i,j)$ making up of subgraphs $j \to t \to i$ for some node $t$
while the denominator counts the number of ordered pairs of nodes $(i,j)$ having no edges connected to node $t$ and only a directed edge
from $j$ to $i$.
Similar to \eqref{thetaabk}, we have
\begin{align}
	\nonumber
	\log \frac{p_{it}^{11}p_{ij}^{10}p_{tj}^{11}}{p_{it}^{10}p_{ij}^{11}p_{tj}^{10}} ~~ = &~~ \big\{
	[(\theta+\alpha_i+\beta_t) + (\theta+\alpha_t+\beta_i) + \rho ] + (\theta+\alpha_i+\beta_j)
	\\ 
	\nonumber
	& ~ + [(\theta+\alpha_t+\beta_j) + (\theta+\alpha_j+\beta_t) + \rho] \big\}
	\\
	\nonumber
	&~ - \big\{ (\theta+\alpha_i+\beta_t) + [(\theta+\alpha_i+\beta_j)+(\theta+\alpha_j+\beta_i)+\rho]
	+(\theta+\alpha_t+\beta_j) \big\}
	\\
	\label{rhoabk}
	=	&~~	\rho+\theta+\alpha_t+\beta_t
\end{align}
for any three distinct nodes $i,j,t\in [n]$.
In view of the constraint in \eqref{constraint}, it follows from \eqref{thetahat} and \eqref{rhoabk} that we can estimate the reciprocity parameter $\rho$ by
\begin{align}\label{rhohat}
	\hat{\rho}=\frac{1}{n}\sum\limits_{t=1}^n\log\frac{\sum_{i,j\neq t}I_{it}^{11}I_{ij}^{10}I_{tj}^{11}}{\sum_{i,j\neq t}I_{it}^{10}I_{ij}^{11}I_{tj}^{10}}-\hat{\theta}.
\end{align}
Further, for five distinct nodes $i,j,k,l,t\in [n]$ we have
\begin{align*}
	\theta+\alpha_i+\beta_t=\log\frac{p_{ki}^{01}p_{kl}^{00}p_{tl}^{01}}{p_{ki}^{00}p_{kl}^{01}p_{tl}^{00}},
\end{align*}	
and
\begin{align*}
	\theta+\alpha_t+\beta_j=\log\frac{p_{kt}^{01}p_{kl}^{00}p_{jl}^{01}}{p_{kt}^{00}p_{kl}^{01}p_{jl}^{00}}.
\end{align*}
Thus, the estimator for $\alpha_i$ is
\begin{align}\label{alphaihat}
	\hat{\alpha}_i=\frac{1}{n}\sum_{t=1}^n\log \frac{\sum_{k,l\neq i,t}I_{ki}^{01}I_{kl}^{00}I_{tl}^{01}}
	{\sum_{k,l\neq i,t}I_{ki}^{00}I_{kl}^{01}I_{tl}^{00}}-\hat{\theta},\quad i\in[n].
\end{align}
Analogously, the estimator of $\beta_j$ is
\begin{align}\label{betajhat}
	\hat{\beta}_j=\frac{1}{n}\sum_{t=1}^n\log \frac{\sum_{k,l\neq j,t}I_{kt}^{01}I_{kl}^{00}I_{jl}^{01}}
	{\sum_{k,l\neq j,t}I_{kt}^{00}I_{kl}^{01}I_{jl}^{00}}-\hat{\theta},\quad j\in[n].
\end{align}
Collecting these estimators above, we denote by
\begin{align*}
	\widehat{\Theta}=(\hat{\rho},\hat{\theta},\hat{\alpha}_1,\ldots,\hat{\alpha}_n,\hat{\beta}_1,\ldots,\hat{\beta}_n)^{\top}
\end{align*}
the estimator of $\Theta$ defined in \eqref{definition-Theta}.
Since our estimators are based on ratios of the sum of a sequence of
triple-dyad indicators $I_{ij}^{ab}I_{kl}^{ab}I_{st}^{ab}$ to another one, we call them \emph{triple-dyad ratio estimators}.

Now, we discuss the computation of the triple-dyad ratio estimators. We use the computation of $\hat{\theta}$ as an illustrating example and other estimators are computed in a similar manner.
Define $I_{ii}^{ab}=0$ for any $a,b\in\{0,1\}$ and let $A^{ab}$ be the matrix made up by elements $\{I_{ij}^{ab}\}_{i,j=1}^n$. The numerator $\sum_{i,j\neq t}I_{it}^{01}I_{ij}^{00}I_{tj}^{01}$ in the expression of $\hat{\theta}$ in \eqref{thetahat} can be written as $\sum_{i,j} I_{tj}^{01} I_{ji}^{00} I_{it}^{01}$, that is, the $t$th diagonal element of $A^{01}A^{00}A^{01}$. Similarly, the denominator $\sum_{i,j\neq t}I_{it}^{00}I_{ij}^{01}I_{tj}^{00}$ in the expression of $\hat{\theta}$ is exactly the $t$th diagonal element of $A^{00}A^{01}A^{00}$. Therefore, the double cyclic calculations for the numerator and denominator can be simplified into a simple algebraic operation. The computational process is summarized in Algorithm \ref{algorithm:a}.

\begin{algorithm}[!htpb]
	\caption{}
	\begin{algorithmic}[1]
		\Require
		Four configuration matrices $A^{00}, A^{01}, A^{10}, A^{11}$
		\Ensure
		Parameter estimates
		\State Compute $B_1=A^{01}A^{00}A^{01}$, $B_2=A^{00}A^{01}A^{00}$, $B_3=A^{11}A^{10}A^{11}$ and $B_4=A^{01}A^{11}A^{01}$
		\For{ $t = 1\to n$ }
		\State $ y_1 = y_1 + (B_1)_{tt}/(B_2)_{tt}$, $y_2 = y_2 + (B_3)_{tt}/(B_4)_{tt}$
		\State $ y_{3,i} = y_{3,i}+(B_1)_{ti}/(B_2)_{it}$, $y_{4,j} = y_{4,j}+ (B_1)_{jt}/(B_2)_{tj}$ 
		\EndFor \\
		\Return $\hat{\theta}=y_1/n$, $\hat{\rho}=(y_2-y_1)/n$, $\hat{\alpha}_i=(y_{3,i}-y_1)/n$ and $\hat{\beta}_j=(y_{4,j}-y_1)/n$
	\end{algorithmic}
	\label{algorithm:a}
\end{algorithm}

\section{Asymptotic properties of estimators}
\label{section-theory}

We now study asymptotic properties of the triple-dyad ratio estimator $\widehat{\Theta}$ as the network size $n$ tends to infinity.
The consistency of all the triple-dyad ratio estimators, $\hat{\rho}, \hat{\theta}, \hat{\alpha}_i$, and $\hat{\beta}_j$, is presented below, and the proof of which is presented in the supplementary material.

\begin{theorem}
	\label{theorem-con}
	(1)	Suppose that $\|\Theta\|_{\infty} \le C$ for a fixed constant $C>0$.
	Then  the triple-dyad ratio estimator $\widehat{\Theta}$ satisfies
	\begin{equation*}
		\|\widehat{\Theta}-\Theta\|_{\infty} = O\bigg( \sqrt{\frac{\log n}{n}} \bigg),   
	\end{equation*}
	with probability at least $1- O(n^{-1})$.
	\\
	(2) Define
	\begin{equation}
		\label{Cn-cn}
		C_n= \max_{i,j}\{p_{ij}^{10}\} = \max_{i,j}\{p_{ij}^{01}\},  \quad c_n= \min_{i,j}\{p_{ij}^{10}\} = \min_{i,j}\{p_{ij}^{01}\}.
	\end{equation}
	If
	\begin{equation}
		\label{con-sparse-a}
		\theta \goto -\infty, \qquad  |\rho| \ll |\theta|.
	\end{equation}
	and
	\begin{equation}
		\label{con-sparse-con}
		\frac{e^{\rho}c_n^{10}}{C_n^8}\gg \frac{\log n}{n},\quad 	\frac{e^{2\rho}c_n^{10}}{C_n^5}\gg \frac{\log n}{n^2},\quad \frac{e^{3\rho/2}c_n^{19/2}}{C_n^{25/4}}\gg\frac{ (\log n)^{3/4} }{n},
	\end{equation}
	hold, then $\widehat{\Theta}$ satisfies
	\begin{eqnarray*}
		\hat{\rho} - \rho  =  
		O\bigg( \sqrt{\frac{C_n^5\log n}{e^{2\rho}n^2c_n^{10}} } + \sqrt{ \frac{C_n^8\log n}{e^{\rho}nc_n^{10}} } \bigg),
	\end{eqnarray*}
	\begin{eqnarray*}
		\max\{ \max_{i\in[n]} | \hat{\alpha}_i - \alpha_i |, \max_{j\in[n]} |\hat{\beta}_j - \beta_j |, |\hat{\theta} - \theta| \}
		=  O\bigg(\sqrt{ \frac{C_n^3\log n}{nc_n^4} } \bigg).
	\end{eqnarray*}
	with probability at least $1- O(n^{-1})$.
\end{theorem}

When $\| \Theta\|_\infty$ is bounded above by a constant,  the network density $\phi_{den}$, defined as $\sum_{i\neq j} \mathbb{E} X_{ij}/[n(n-1)]$,
is of a constant order,  
which implies that the total number of edges is proportional to $n^2$.
The condition $\theta \goto -\infty$ 
is widely used to describe sparse networks \cite[e.g.][]{Wang2017aos,Chen:2020}. The other condition $|\rho| \ll |\theta|$ means that
the reciprocity effect is much smaller than the effect of sparsity. This is a mild condition since $\theta$ is a global density parameter.

\begin{remark}
	Condition \eqref{con-sparse-a} implies
	\[
	e^{\rho}c_n^2 \leq p_{ij}^{11} \leq e^{\rho}C_n^2
	\]
	for any $i\neq j$.
	Condition \eqref{con-sparse-con} is a technical condition,
	restricting the increasing rate of $C_n$ and the decreasing rate of $c_n$.
	If
	$\rho$, $\max_i |\alpha_i|$, $\max_i|\beta_i|$ are bounded above by a constant,
	then Conditions \eqref{con-sparse-a} and \eqref{con-sparse-con} imply
	$p_{ij}^{00} \approx 1$, $C_n\asymp e^\theta$ and $c_n \asymp e^\theta$.
	In this case,  the network density $\phi_{den}$
	is of the order $O(e^{\theta})$
	and condition \eqref{con-sparse-con} becomes
	$\phi_{den} \gg (\log n)^{1/5}/n^{2/5} $.
	That is, the allowed smallest network density is in the order of $O( n^{-2/5}/\log n ) $, up to a logarithmic factor.
\end{remark}

\begin{remark}
	In Theorem \ref{theorem-con}, the consistency rates for all the estimators $\hat{\rho}, \hat{\theta}, \hat{\alpha}_i$ and $\hat{\beta}_j$ are the same order of $(\log n/n)^{1/2}$ if $\Theta$ is a constant. On the one hand, $\hat{\rho}$ and $\hat{\theta}$ may have faster convergence rates
	since we observe $O(n^2)$ observed dyads while there exists only one parameter each for the density and reciprocity.
\end{remark}

We now derive the asymptotic distribution of the proposed triple-dyad ratio estimators. To describe our results, we start by introducing the notation as follows.
For any $a,b,c \in \{0,1\}$, we define
\begin{eqnarray}
	\label{mutabc}
	\mu_{t}^{(abc)}=\frac{1}{n^2}\sum\limits_{i,j\neq t}p_{it}^{ca}p_{ij}^{cb}p_{tj}^{ca}, \quad t\in[n],
	\\
	\label{muitabc}
	\mu_{it}^{(abc)}=\frac{1}{n^2}\sum\limits_{k,l\neq i,t}p_{ki}^{ca}p_{kl}^{cb}p_{tl}^{ca}, \quad 1\le i\neq t\le n.
\end{eqnarray}
For any $1\le i\neq t\le n$, define
\begin{align}
	\label{etait}
	\eta_{it}^{(abc)}&=\frac{1}{n}\sum\limits_{j\neq i,t}\left(\frac{1}{\mu_{t}^{(abc)}}p_{tj}^{ca}p_{ij}^{cb}+\frac{1}{\mu_{i}^{(abc)}}p_{ji}^{ca}p_{jt}^{cb}
	-\frac{1}{\mu_j^{(bac)}}p_{ij}^{cb}p_{jt}^{cb}\right), \\
	\zeta_{it,1}^{(abc)}&=\frac{1}{2n^3}\bigg[\bigg(\frac{1}{\mu_t^{(abc)}}\sum\limits_{j\neq t}p_{tj}^{ca}p_{ij}^{cb}\bigg)^2+\bigg(\frac{1}{\mu_i^{(abc)}}\sum\limits_{j\neq t}p_{ji}^{ca}p_{jt}^{cb}\bigg)^2\bigg],\notag\\
	\label{zetait}
	\zeta_{it,2}^{(abc)}&=\frac{1}{n^3}\bigg(\frac{1}{\mu_t^{(abc)}}\sum\limits_{j\neq t}p_{tj}^{ca}p_{ij}^{cb}\bigg)\bigg(\frac{1}{\mu_t^{(abc)}}\sum\limits_{j\neq t}p_{jt}^{ca}p_{ji}^{cb}\bigg).
\end{align}
For any distinct nodes $i,j,k,l\in [n]$, we define
\begin{align}\label{kappaik}
	&\kappa_{ik}^{(1)}=\frac{1}{n^2}\sum\limits_{t\neq i,k}\sum\limits_{l\neq i,k,t}\frac{1}{\mu_{it}^{(100)}}p_{tl}^{01}p_{kl}^{00},\quad
	\kappa_{ik}^{(2)}=\frac{1}{n^2}\sum\limits_{t\neq i,k}\sum\limits_{l\neq i,k,t}\frac{1}{\mu_{it}^{(010)}}p_{tl}^{00}p_{kl}^{01},
\end{align}
\begin{align}\label{xijl}
	&\xi_{jl}^{(1)}=\frac{1}{n^2}\sum\limits_{t\neq j,l}\sum\limits_{k\neq j,l,t}\frac{1}{\mu_{tj}^{(100)}}p_{kt}^{01}p_{kl}^{00},\quad
	\xi_{jl}^{(2)}=\frac{1}{n^2}\sum\limits_{t\neq j,l}\sum\limits_{k\neq j,l,t}\frac{1}{\mu_{tj}^{(010)}}p_{kt}^{00}p_{kl}^{01}.
\end{align}
Under the assumption that $\|\Theta\|\le C$ for some constant $C>0$,
we can obtain that $\mu_{t}^{(abc)}, \mu_{it}^{(abc)}, \eta_{it}^{(abc)}\asymp 1$ for any $a,b,c=0,1$, and $\kappa_{ik}^{(1)},\kappa_{ik}^{(2)},\xi_{jl}^{(1)},\xi_{jl}^{(2)}\asymp 1$ for any distinct $i,j,k,l\in [n]$.

For any positive integer $m$ and two real vectors $(x_1,\ldots,x_m)^{\top}$ and $(y_1,\ldots,y_m)^{\top}$, we define a $2m$-variate function
\begin{align}
	\label{def:g_k}
	g_m(x_1, \ldots, x_m; y_1, \ldots, y_m)=x_1^2y_1 + \cdots +  x_m^2y_m - (x_1y_1+ \cdots + x_my_m)^2.
\end{align}
Define
\begin{align}\label{gammaTheta}
	\left\{\begin{array}{cl}
		\sigma_{\theta}^2&=\frac{1}{n^4}\sum\limits_{t<i}g_3\Big(\eta_{it}^{(100)},\eta_{ti}^{(100)},
		-\big(\eta_{it}^{(010)}+\eta_{ti}^{(010)}\big);~p_{it}^{01},p_{it}^{10},p_{it}^{00}\Big),\\
		\sigma^2_{\rho}&=\frac{1}{n^4}\sum\limits_{t<i}g_4\Big(-\big(\eta_{it}^{(100)}+\eta_{ti}^{(011)}\big),\eta_{it}^{(010)}+\eta_{ti}^{(010)},\eta_{it}^{(101)}+\eta_{ti}^{(101)},\\
		&\quad-\big(\eta_{it}^{(011)}+\eta_{ti}^{(100)}\big);~p_{it}^{01},p_{it}^{00},p_{it}^{11},p_{it}^{10}\Big),\\
		\sigma_{\alpha_i}^2&=\frac{1}{n^2}\sum\limits_{k\neq i}g_2\left(\kappa_{ik}^{(1)},-\kappa_{ik}^{(2)};
		~p_{ki}^{01},p_{ki}^{00}\right),\quad i\in [n],\\
		\sigma_{\beta_j}^2&=\frac{1}{n^2}\sum\limits_{l\neq j}g_2\left(\xi_{jl}^{(1)},
		-\xi_{jl}^{(2)};~p_{jt}^{01},p_{jt}^{00}\right),\quad j\in [n],\\
		\sigma_{ii}&=\frac{1}{n^2}\sum\limits_{l\neq i}\Big(-\kappa_{il}^{(1)}\xi_{il}^{(1)}p_{il}^{10}p_{il}^{01}+\kappa_{il}^{(1)}\xi_{il}^{(2)}p_{il}^{10}p_{il}^{00}+\kappa_{il}^{(2)}\xi_{il}^{(1)}p_{il}^{01}p_{il}^{01}\\
		&\quad+\kappa_{il}^{(2)}\xi_{il}^{(2)}p_{il}^{00}(1-p_{il}^{01})\Big),\quad i\in [n].
	\end{array}\right.
\end{align}
Similarly, under the condition that $\|\Theta\|_\infty \le C$, we also have that $\sigma_{\theta}^2,\sigma_{\rho}^2,\sigma_{\theta,\rho}\asymp n^{-2}$ and $\sigma_{\alpha_i}^2,\sigma_{\beta_j}^2,\sigma_{ii}\asymp n^{-1}$.
We further denote
\begin{align}\label{def:theta*} \theta^*&=\frac{1}{n^2}\sum\limits_{t<i}\big[-\zeta_{it,1}^{(100)}p_{it}^{01}\big(1-p_{it}^{01}\big)-\zeta_{ti,1}^{(100)}p_{it}^{10}\big(1-p_{it}^{10}\big)+\big(\zeta_{it,2}^{(100)}+\zeta_{ti,2}^{(100)}\big)p_{it}^{10}p_{it}^{01}\notag\\
	&\quad+\big(\zeta_{it,1}^{(010)}+\zeta_{ti,1}^{(010)}+\zeta_{it,2}^{(010)}+\zeta_{ti,2}^{(010)}\big)p_{it}^{00}\big(1-p_{it}^{00}\big)\big],
\end{align}
and
\begin{align}\label{def:rho^*} \rho^*&=\frac{1}{n^2}\sum\limits_{t<i}\big(\zeta_{it,1}^{(100)}+\zeta_{ti,1}^{(011)}\big)p_{it}^{01}\big(1-p_{it}^{01}\big)+\big(\zeta_{ti,1}^{(100)}+\zeta_{it,1}^{(011)}\big)p_{it}^{10}\big(1-p_{it}^{10}\big)\notag\\
	&\quad-\big(\zeta_{it,1}^{(101)}+\zeta_{ti,1}^{(101)}+\zeta_{it,2}^{(101)}+\zeta_{ti,2}^{(101)}\big)p_{it}^{11}\big(1-p_{it}^{11}\big)\notag\\
	&\quad-\big(\zeta_{it,1}^{(010)}+\zeta_{ti,1}^{(010)}+\zeta_{it,2}^{(010)}+\zeta_{ti,2}^{(010)}\big)p_{it}^{00}\big(1-p_{it}^{00}\big)\notag\\
	&\quad-\big(\zeta_{it,2}^{(100)}+\zeta_{ti,2}^{(100)}-\zeta_{it,2}^{(011)}-\zeta_{ti,2}^{(011)}\big)p_{it}^{01}p_{it}^{10}.
\end{align}	
After straightforward calculations with \eqref{def:theta*} and \eqref{def:rho^*}, we obtain that both $\theta^*$ and $\rho^*$ are, at most, of order $n^{-1}$ if $\|\Theta\|_\infty \le C$ for some constant $C>0$.

We can now present the central limit theorem of the triple-dyad ratio estimator.

\begin{theorem}\label{theorem-central}
	Suppose that $\|\Theta\|_{\infty} \le C$ for a fixed constant $C>0$.
	Then as $n\rightarrow \infty$, we have
	\begin{eqnarray}
		\label{proof-clt-theta}
		\frac{\hat{\theta}-\theta-\theta^*}{\sigma_{\theta}}\stackrel{d}{\longrightarrow}\mathcal{N}(0,1),
		\\
		\label{proof-clt-rho}
		\frac{\hat{\rho}-\rho-\rho^*}{\sigma_{\rho}}\stackrel{d}{\longrightarrow}\mathcal{N}(0,1),
	\end{eqnarray}
	where $\stackrel{d}{\longrightarrow}$ denotes ``convergence in distribution,'' and for any fixed positive integers $k_1$ and $k_2$,
	\begin{align}\label{proof-clt-alpha-beta}
		\bm{\Sigma}^{-\frac{1}{2}}\left(\hat{\alpha}_{1}-\alpha_1,\ldots,\hat{\alpha}_{k_1}-\alpha_{k_1},
		\hat{\beta}_{1}-\beta_1,\ldots,\hat{\beta}_{k_2}-\beta_{k_2}
		\right)^{\top}\stackrel{d}{\longrightarrow} \mathcal{N}(0,\bm I_{k_1+k_2}),
	\end{align}
	where
	\[
	\bm{\Sigma}=\left(\begin{array}{cc}
		\bm{\Sigma}_{11} & \bm{\Sigma}_{12} \\
		\bm{\Sigma}_{12} & \bm{\Sigma}_{22}
	\end{array}\right)
	\]
	and $\bm{\Sigma}_{11}={\rm diag}(\sigma_{\alpha_1}^2,\ldots,\sigma_{ \alpha_{k_1} }^2)$,
	$\bm{\Sigma}_{22}={\rm diag}(\sigma_{\beta_1}^2,\ldots,\sigma_{\beta_{k_2}}^2)$ and
	$\bm{\Sigma}_{12}$ is a $k_1\times k_2$ matrix with the $(i,i)$-entry $\sigma_{ii}$ for $i=1,\ldots, \min\{k_1,k_2\}$ and other entries $0$.
	In addition, if \eqref{con-sparse-a} holds, and $(C_n,c_n)$ defined in \eqref{Cn-cn} satisfy
	\begin{equation}
		\label{eq-sparse-clt}
		\frac{c_n^{15}}{C_n^{13}}\gg \frac{\log^3 n}{n},\quad \frac{e^{\rho}c_n^8}{C_n^5}\gg \frac{\log n}{n}
	\end{equation}
	then \eqref{proof-clt-theta}, \eqref{proof-clt-rho} and \eqref{proof-clt-alpha-beta} continue to hold.
\end{theorem}

The proof of Theorem \ref{theorem-central} is given in the supplementary material.
The calculations of the variances of the triple-dyads estimators are very complex
since we need to analyze hundreds of terms that come from divisions of their second-order Taylor's expansions.
Furthermore, the squared terms in Taylor's expansion of $\hat{\theta} - \theta$ and $\hat{\rho}-\rho$ are not neglected and their orders need to be computed,
which leads to a bias in their asymptotic distributions.
In addition, the number of the leading terms are not one or two. We need to compute the variance of each term.
In addition, the terms are not independent such that we need to analyze many different cases for the triple dyads.

\begin{remark}
	The asymptotic distributions of $\hat{\rho}$ and $\hat{\theta}$ contain bias terms $\rho^*$ and $\theta^*$ defined in \eqref{def:theta*}
	and \eqref{def:rho^*}, respectively.
	In contrast, the asymptotic distributions of  $\{\hat{\alpha}_i\}_{i}$ and $\{\hat{\beta}_j\}_{j}$ do not contain bias terms.
	Bias-corrected procedures are needed to valid inference for $\hat{\rho}$ and $\hat{\theta}$; see next section.
\end{remark}

\begin{remark}
	If all parameters are of constant orders, then the asymptotic variances of $\hat{\rho}$ and $\hat{\theta}$ are of order $1/n^{2}$, while those of $\{\hat{\alpha}_i\}_{i\in[n]}$ and $\{\hat{\beta}_j\}_{j\in[n]}$ are of order $1/n$.
	If $\rho$, $\max_i |\alpha_i|$ and $\max_i|\beta_i|$ are bounded above by a constant, except for $\theta$,
	then $C_n\asymp e^\theta$ and $c_n \asymp e^\theta$ such that condition \eqref{eq-sparse-clt} becomes
	$e^{\theta} \gg (\log n)^{1/3}/n^{1/3} $.
	In this case, the allowed smallest network density is in the order of $O( n^{-1/3}\log n ) $, up to a logarithm factor.
	The asymptotic variances of $\hat{\rho}$ and $\hat{\theta}$ are of order $n^{-2}e^{-2\theta}$ and $n^{-2}e^{-\theta}$, while those of $\{\hat{\alpha}_i\}_{i\in[n]}$ and $\{\hat{\beta}_j\}_{j\in[n]}$ are of orders $\kappa_i(\sum_i \omega_i)$ and $\omega_j(\sum_i \kappa_i)$ respectively, where $\kappa_i=e^{\theta/2+\alpha_i}$ and $\omega_j=e^{\theta/2+\beta_j}$.
\end{remark}

\section{Testing the reciprocal effect}
\label{section-test}
Although Theorem \ref{theorem-central} presents the asymptotic distribution of $\hat{\rho}$, it cannot be directly used to construct a test statistic for testing the reciprocal effect $H_0:\rho=0$. This is because the asymptotic distribution contains a bias term $\rho^*$ and unknown variance $\sigma_{\rho}^2$, the expressions of which are given in \eqref{gammaTheta} and \eqref{def:rho^*}, respectively. It is thus natural to use the plug-in estimates for $\rho^*$ and $\sigma_{\rho}^2$, denoted by $\hat{\sigma}_{\rho}^2$ and $\hat{\rho}^*$, where the unknown parameters are replaced by their triple-dyad ratio estimates.

The following lemma presents consistency of $\hat{\sigma}_{\rho}^2$ and
the error bound between $\hat{\rho}^*$ and $\rho^*$.

\begin{lemma}\label{pro-est}
	
	Suppose that $\|\Theta\|_{\infty}\leq C$ for a fixed constant $C>0$. Then, as $n\rightarrow\infty$, we have
	\begin{align*}
		\frac{\hat{\sigma}_{\rho}^2}{{\sigma}_{\rho}^2}\stackrel{p}{\longrightarrow} 1,
		\quad\text{and}\quad
		\hat{\rho}^*-\rho^*=O\bigg(\sqrt{\frac{\log n}{n^3}}\bigg)
	\end{align*}
	with probability at least $1-O(n^{-1})$, where $\stackrel{p}{\longrightarrow}$ denotes ``convergence in probability.''
\end{lemma}

From the above lemma, we know that $\hat{\rho}^*-\rho^*$ has a faster convergence rate than the order $O(n^{-1})$ of the standard error $\sigma_\rho$. Therefore, using Slutsky's theorem, we have the following corollary:

\begin{corollary}
	\label{coro-reciprocal}
	Suppose that $\|\Theta\|_{\infty}\leq C$ for a fixed constant $C>0$. Then, as $n\rightarrow\infty$, we have
	\begin{align}\label{asy-est-rho}
		\frac{\hat{\rho}-\rho^*-\hat{\rho}^*}{\hat{\sigma}_{\rho}}\stackrel{d}{\longrightarrow}\mathcal{N}(0,1).
	\end{align}
\end{corollary}

By Corollary \ref{coro-reciprocal}, we can test for the reciprocal effect, that is, whether $H_0:\rho=0$ by using the test statistic
$|\hat{\rho}-\hat{\rho}^*|/\hat{\sigma}_{\rho}$. Under the test level $\alpha$, the null is rejected if $|\hat{\rho}-\hat{\rho}^*|>z_{1-\alpha/2}\hat{\sigma}_{\rho}$,
where $z_{\alpha}$ denotes the $100\alpha$ percentile point of
the standard normal distribution. 
An approximate $100(1-\alpha)\%$ confidence interval for $\rho^*$ is $\hat{\rho}-\hat{\rho}^* \pm z_{1-\alpha/2}\hat{\sigma}_{\rho}$.
Another potential application is to compare whether two different graphs have the same reciprocity effect.
Let $G_1$ and $G_2$ be two independent graphs, where the corresponding estimates are denoted by $\hat{\rho}_i$, 
$\rho^*_i$, $\hat{\rho}^*_i$  and $\hat{\sigma}_{\rho,i}$, $i=1,2$.
For testing the null $\rho^*_1=\rho_2^*$, we can construct the test statistic 
$T(G_1,G_2)=\{ (\hat{\rho}_1 - \hat{\rho}^*_1) - (\hat{\rho}_2 - \hat{\rho}^*_2) \}/( \hat{\sigma}_{\rho,1}^2 + \hat{\sigma}_{\rho,2}^2 )^{1/2}$
and reject the null if $|T(G_1,G_2)|>z_{1-\alpha/2}$ at the test level $\alpha$.

\begin{remark}
	According to Theorems \ref{theorem-con} and \ref{theorem-central},
	an approximate $100(1-\alpha)\%$ confidence interval for $\alpha_i-\alpha_j$ is $\hat{\alpha}_i - \hat{\alpha}_j \pm z_{1-\alpha/2}
	(\hat{\sigma}_{\alpha_i}^2+\hat{\sigma}_{\alpha_j}^2)^{1/2}$, where
	$\hat{\sigma}_{\alpha_i}$ is the plug-in estimate of $\sigma_{\alpha_i}$.
	To test whether $\alpha_i=\alpha_j$ at level $\alpha$, the hypothesis can be rejected if $|\hat{\alpha}_i-\hat{\alpha}_j|> z_{1-\alpha/2}(\hat{\sigma}_{\alpha_i}^2+\hat{\sigma}_{\alpha_j}^2)^{1/2}$.
	Similarly, we can construct Wald-type test statistics for testing the equality of several parameters. For example, a test statistic for the null $\alpha_1=\alpha_2=\alpha_3=\alpha_4$ is
	\[
	(\hat{\alpha}_{1} -\hat{\alpha}_{2}, \hat{\alpha}_{2} -\hat{\alpha}_{3}, \hat{\alpha}_{3}-\hat{\alpha}_{4})
	\begin{pmatrix}
		\hat{\sigma}_{\alpha_{1}}^2+\hat{\sigma}_{\alpha_{2}}^2 & -\hat{\sigma}_{\alpha_{2}}^2 & 0 \\
		-\hat{\sigma}_{\alpha_{2}}^2 & \hat{\sigma}_{\alpha_{2}}^2+\hat{\sigma}_{\alpha_{3}}^2 & \hat{\sigma}_{\alpha_{3}}^2 \\
		0 & \hat{\sigma}_{\alpha_{3}}^2 & \hat{\sigma}_{\alpha_3}^2+\hat{\sigma}_{\alpha_{4}}^2
	\end{pmatrix}^{-1}
	\begin{pmatrix}
		\hat{\alpha}_1 -\hat{\alpha}_2 \\ \hat{\alpha}_2 -\hat{\alpha}_3 \\ \hat{\alpha}_3-\hat{\alpha}_4
	\end{pmatrix},
	\]
	which asymptotically follows the chi-square distribution with $3$ degrees of freedom.
\end{remark}

\begin{remark}
	We can also use the plug-in estimate $\hat{\theta}^*$ to estimate the unknown bias $\theta^*$ in \eqref{proof-clt-theta}. Then, similar to Corollary \ref{coro-reciprocal}, we have $|\hat{\theta}^*-\theta^*|=O_p( (\log n)^{1/2}/n^{3/2})$, and $(\hat{\theta}-\theta - \hat{\theta}^*)/\hat{\sigma}_\theta$ converges to the standard normal distribution. Thus, an approximate $100(1-\alpha)\%$ confidence interval for $\theta$ is 
	$\hat{\theta}-\hat{\theta}^* \pm z_{1-\alpha/2}\hat{\sigma}_{\theta}$.
\end{remark}

\section{Numerical experiments} \label{section-simulation}

In this section, we evaluate the performance of the proposed triple-dyad ratio estimator in networks of finite sizes.

\subsection{Simulations}
The parameters are set as follows. We set the reciprocity parameter $\rho=0.5$, a positive signal for the effect of mutual edges.
The parameters $\alpha_i$ and $\beta_j$ are specified via a linear type as
\[ \beta_i = \alpha_i =
\begin{cases} i/(n/2), & i = 1, \ldots, n/2 \\
	-(i - n/2)/(n/2),   & i = n/2+1, \ldots, n,
\end{cases}
\]
This case was considered in \cite{Yan2015p1} and \cite{Yan2016exponential}.
The design satisfies the model identification given in \eqref{constraint}.

First, we evaluate the estimation error for all the parameters and compare them with the error of the MLE.
We use the frequency iterative algorithm proposed by \cite{Holland1981exponential} to solve the MLE.
The network sizes are set to be $n= 500, 1000, 5000$.
We choose four different density parameters, $\theta$, to evaluate the different asymptotic regimes, that is, $\theta=-(\log n)/3, -(\log n)/4, -\log(\log n)$ and $0$.
(We try a smaller value for $\theta$, i.e., $\theta=-(\log n)/2$ and find that the MLE failed to exist in all repeated simulations.)
Each simulation is repeated $1000$ times, except for $n=5000$, which we only repeat $100$ times because computing the MLE for a large $n$ would be too time consuming.

We record the average values of the absolute errors for the triple-dyad ratio estimators and MLEs,
that is, $|\hat{\theta}-\theta|$ ($|\tilde{\theta}-\theta|)$, $|\hat{\rho} - \rho|$ ($|\tilde{\rho} - \rho|$),
$\| \hat{\alpha} - \alpha \|_\infty$ ($\| \tilde{\alpha} - \alpha \|_\infty$), 
$|\hat{\alpha}_i - \alpha_i|$ ($|\tilde{\alpha}_i - \alpha_i|$) for several $i$ values, where the symbol $\tilde{~~}$ denotes the value of the MLE.
Table \ref{table-a} presents the simulation results; as the errors for $\beta_j$ are similar to $\alpha_j$, so we do not represent them to save space.

\begin{table}[h!]\centering
	\caption{ The estimation errors for triple-dyad ratio estimator and MLE. 
	The symbols $\hat{~}$ and $\tilde{~}$ denote the triple-dyad ratio estimator and MLE, respectively. }
	\label{table-a}
	\small
	\renewcommand\arraystretch{1.5}
	\begin{threeparttable}
		\begin{tabular}{llll l}
			\hline
			& $\theta=-(\log n)/3$     & $\theta=-(\log n)/4$ & $\theta=-\log(\log n)$ & $\theta=0$ \\
			\hline
			&	\multicolumn{4}{c}{ $n=500$} \\
			\hline
			$|\hat{\theta} - \theta|/|\tilde{\theta} - \theta|$   &   $ 0.017 / 0.016 $ & $ 0.01 / 0.011  $&$ 0.013 / 0.013  $&$ 0.01 / 0.008  $ \\
			$|\hat{\rho} - \rho|/|\tilde{\rho}-\rho|$   &  $ 0.137 / 0.015 $   & $ 0.06 / 0.013  $&$ 0.035 / 0.014  $&$ 0.013 / 0.014  $ \\
			$|\hat{\alpha} - \alpha |_{m}/ |\tilde{\alpha} - \alpha|_{m} $ & $ 0.618 / 0.548 $ & $ 0.528 / 0.451  $&$ 0.574 / 0.509  $&$ 0.562 / 0.326  $ \\
			$|\hat{\alpha}_1 - \alpha_1|/|\tilde{\alpha}_1 - \alpha_1|$  & $ 0.13 / 0.118 $ & $ 0.103 / 0.089  $&$ 0.113 / 0.096  $&$ 0.109 / 0.074  $
			\\
			$|\hat{\alpha}_{\tfrac{n}{2}} - \alpha_{\tfrac{n}{2}}|/|\tilde{\alpha}_{\tfrac{n}{2}} - \alpha_{\tfrac{n}{2}}|$ & $ 0.109 / 0.089 $ & $ 0.104 / 0.075  $
			&$ 0.104 / 0.08  $&$ 0.167 / 0.092  $ \\
			$|\hat{\alpha}_{n} - \alpha_{n}|/|\tilde{\alpha}_n - \alpha_n|$& $ 0.185 / 0.175 $ & $ 0.142 / 0.127  $&$ 0.155 / 0.145  $&$ 0.105 / 0.078  $ \\
			\hline
			& \multicolumn{4}{c}{ $n=1000$ } \\
			\hline
			$|\hat{\theta} - \theta|/|\tilde{\theta} - \theta|$     & $ 0.01 / 0.01 $ & $ 0.006 / 0.006  $&$ 0.007 / 0.007  $&$ 0.005 / 0.007  $ \\
			$|\hat{\rho} - \rho|/|\tilde{\rho}-\rho|    $      & $ 0.042 / 0.01 $  & $ 0.046 / 0.008  $&$ 0.064 / 0.008  $&$ 0.007 / 0.01  $ \\
			$|\hat{\alpha} - \alpha |_{m}/ |\tilde{\alpha} - \alpha|_{m} $
			& $ 0.518 / 0.484 $ & $ 0.406 / 0.361  $&$ 0.442 / 0.396  $&$ 0.418 / 0.242  $ \\
			$|\hat{\alpha}_1 - \alpha_1|/|\tilde{\alpha}_1 - \alpha_1|$  & $ 0.09 / 0.084 $ & $ 0.08 / 0.068  $&$ 0.08 / 0.071  $&$ 0.081 / 0.054  $ \\
			$|\hat{\alpha}_{\tfrac{n}{2}} - \alpha_{\tfrac{n}{2}}|/|\tilde{\alpha}_{\tfrac{n}{2}} - \alpha_{\tfrac{n}{2}}|$ &  $ 0.074 / 0.065 $
			& $ 0.074 / 0.057  $ & $ 0.074 / 0.057  $ &
			$ 0.12 / 0.06  $ \\
			$|\hat{\alpha}_{n} - \alpha_{n}|/|\tilde{\alpha}_n - \alpha_n|$ & $ 0.127 / 0.12 $ & $ 0.109 / 0.104  $&$ 0.113 / 0.107  $&$ 0.075 / 0.059  $ \\
			\hline
			&	\multicolumn{4}{c}{ $n=5000$ } \\
			\hline
			$|\hat{\theta} - \theta|/|\tilde{\theta} - \theta|$  &  $ 0.004 / 0.003 $  & $ 0.002 / 0.002  $&$ 0.001 / 0.027  $&$ 0.001 / 0.001  $ \\
			$|\hat{\rho} - \rho|/|\tilde{\rho}-\rho|   $   &  $ 0.055 / 0.003$     & $ 0.019 / 0.002  $&$ 0.002 / 0.043  $&$ 0.002 / 0.001  $ \\
			$|\hat{\alpha} - \alpha |_{m}/ |\tilde{\alpha} - \alpha|_{m} $ & $ 0.318 / 0.306$ & $ 0.23 / 0.214  $&$ 0.256 / 0.168  $&$ 0.153 / 0.106  $ \\
			$|\hat{\alpha}_1 - \alpha_1|/|\tilde{\alpha}_1 - \alpha_1|$   & $ 0.043 / 0.042$ & $ 0.034 / 0.032  $&$ 0.04 / 0.028  $&$ 0.028 / 0.02  $ \\
			$|\hat{\alpha}_{\tfrac{n}{2}} - \alpha_{\tfrac{n}{2}}|/|\tilde{\alpha}_{\tfrac{n}{2}} - \alpha_{\tfrac{n}{2}}|$ & $ 0.040 / 0.032$
			& $ 0.033 / 0.025  $&$ 0.044 / 0.048  $&$ 0.031 / 0.021  $ \\
			$|\hat{\alpha}_{n} - \alpha_{n}|/|\tilde{\alpha}_n - \alpha_n|$ & $ 0.072 / 0.068$ & $ 0.055 / 0.055  $&$ 0.067 / 0.058  $&$ 0.029 / 0.019  $ \\
			\hline
		\end{tabular}
		
		\begin{tablenotes}
			\footnotesize
			\item $|\hat{\alpha} - \alpha |_{m}$ and $|\tilde{\alpha} - \alpha|_{m}$ mean
			$\max_i |\hat{\alpha}_i - \alpha_i |$ and $\max_i |\tilde{\alpha}_i -\alpha_i |$, respectively.
		\end{tablenotes}
		
	\end{threeparttable}
\end{table}

\begin{table}[htb!]
	\centering
	\scriptsize
	\caption{Computing time for our method (TRE) and MLE in seconds (average in 10 repetitions). In many cases, MLE is $10$ times slower than our method ($\rho = 0.5$).}
	\label{table-compare-time}
	\renewcommand\arraystretch{1.5}
	\begin{tabular}{c ccc ccc ccc ccc cc}
		\hline
		$n$   & \multicolumn{2}{c}{$\theta=-\tfrac{1}{3}\log n$} & & \multicolumn{2}{c}{$\theta=-\tfrac{1}{6}\log n$} & & \multicolumn{2}{c}{$\theta=-\log(\log n)$}
		& & \multicolumn{2}{c}{$\theta=0$}  & &  \multicolumn{2}{c}{$\theta=0.5 $}      \\
		\hline
		& TRE  & MLE  & & TRE & MLE  & & TRE & MLE   && TRE & MLE  & & TRE & MLE \\
		\cline{2-3} \cline{5-6}  \cline{8-9} \cline{11-12} \cline{14-15}
		$500$  & $0.858$&$  1.864$ && $0.864$&$  2.781$ && $0.874$&$  2.244$ && $0.836$&$  4.889$ && $0.859$&$  6.591$ \\
		$1000$ & $1.014$ &  $7.824$ && $1.065$&  $12.526$ && $1.046$& $8.767$ && $1$&  $20.526$ && $0.987$& $26.233$ \\
		$2000$ & $1.419$&  $24.752$ && $1.479$&  $39.254$ && $1.451$&  $29.123$ && $1.45$&  $66.808$ && $1.502$& $87.915$\\
		\hline
	\end{tabular}
\end{table}

Table \ref{table-a} shows that the errors of our proposed  estimator $\hat{\theta}$ are comparable to those of the MLE $\tilde{\theta}$ when $n\ge 1000$.
Especially,  when $n$ is very large, (e.g., $n=5000$), the average value of $|\tilde{\theta}-\theta|$  and $|\hat{\theta}-\theta|$ are very close.
The same phenomenon can be observed for $\hat{\rho}$. This indicates a very high accuracy for our proposed estimators. On the other hand, the error of the estimator $\hat{\alpha}_i$ or $\|\hat{\alpha}\|_\infty$ is larger than the corresponding MLE but not more than twice, when $n\le 1000$. When $n$ increases to $5000$, the difference between $\hat{\alpha}_i$ and $\tilde{\alpha}_i$ is very close, up to two decimal places.

We also compare the average running time between our estimate and the MLE on an Intel T7700 2.40GHz machine with 16GB memory,
shown in Table \ref{table-compare-time}.
From this table, we can see that the computing time for our estimator is much faster than the MLE.
This is due to that our proposed estimators have explicit expressions while the computation of the MLE needs to an iterative algorithm.
In particular, when $n=2000$, the difference is up to twenty times.

Next, we evaluate asymptotic normality of the triple-dyad-ratio estimator in Theorem \ref{theorem-central}. For simplicity, we consider only one network size $n=1000$ and choose three density parameters, $\theta=-0.5, 0, 0.5$. Each simulation is repeated $5,000$ times. We record the values of $(\hat{\theta} - \theta)/\hat{\sigma}_\theta$, $(\hat{\rho} - \rho)/\hat{\sigma}_\rho$, $ ( \hat{\alpha}_i - \alpha_i )/\hat{\sigma}_{\alpha_i}$, and $(\hat{\beta}_i - \beta_i)/ \hat{\sigma}_{\beta_i}$, and then draw their quantile--quantile plots to assess asymptotically normal approximation---where $\hat{\sigma}^2_\theta$, $\hat{\sigma}^2_\rho$, $\hat{\sigma}^2_{\alpha_i}$,
and $\hat{\sigma}^2_{\beta_i}$ are the respective estimators of the asymptotic variances (given in Theorem \ref{theorem-central})
of $\hat{\theta}$, $\hat{\rho}$, $\hat{\alpha}_i$, and $\hat{\beta}_j$---by replacing their unknown parameters with their estimators.
We also reported the $95\%$ coverage probabilities for $\theta$, $\rho$, $\alpha_i$, and $\beta_j$, as shown in Table \ref{Table:b}.

\begin{figure}[!htp]
	\centering
	\includegraphics[height=2.38in, width=4.05in, angle=0]{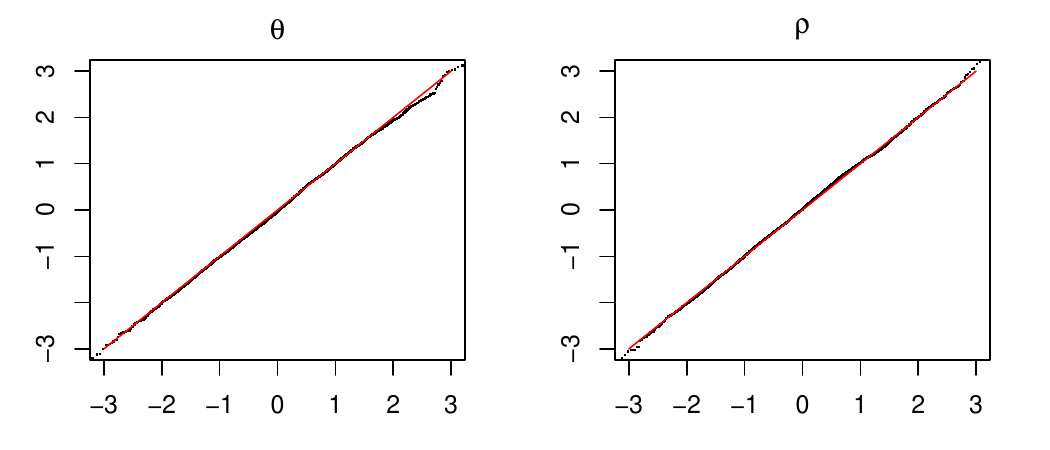}
	\caption{QQ-plots for $(\hat{\theta}-\theta-\theta^*)/\sigma_{\theta}$ and $(\hat{\rho}-\rho-\rho^*)/\sigma_{\rho}$.
		The horizontal and vertical axes are the theoretical and sample quantiles. The red color indicates the diagonal line. }
	\label{fig-rho-theta}
\end{figure}

\begin{figure}[!htp]
	\centering
	\includegraphics[height=3.68in,width=5.55in, angle=0]{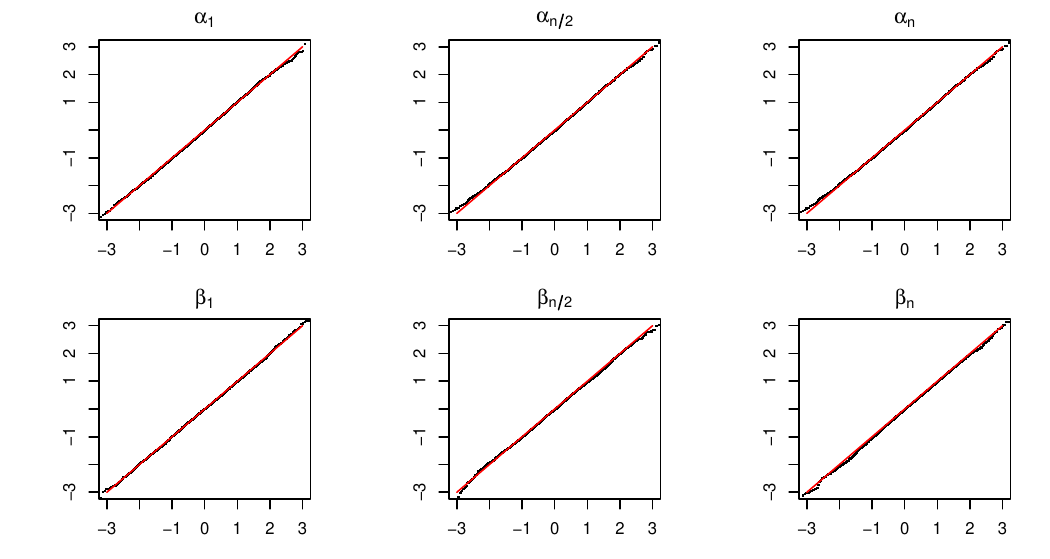}
	\caption{QQ-plots for $(\hat{\alpha}_i-\alpha_i)/\sigma_{\alpha_i}$ and $(\hat{\beta}_i-\beta_i)/\sigma_{\beta_i}$, $i=1,n/2,n$. The horizontal and vertical axes are the theoretical and sample quantiles. The red color indicates the diagonal line.}\label{fig-alpha-beta}
\end{figure}

Figures \ref{fig-rho-theta} and \ref{fig-alpha-beta} show that the sample quantiles agree with the quantiles of the standard normal distribution very well, indicating that the approximation of asymptotic normality in Theorem \ref{theorem-central} is good when all parameters are bounded above by a constant. Table \ref{Table:b} further shows that all simulated coverage frequencies are very close to the target level.

\begin{table}[h]\centering
	\caption{Coverage frequencies ($\times 100$) of $95\%$ confidence intervals} 
	\label{Table:b}
	%\scriptsize
	\vskip5pt
	\begin{tabular}{ccc ccc ccc}
		\hline
		& $\rho$ & $\theta$ & $\alpha_1$ & $\alpha_{n/2}$ & $\alpha_{n}$ & $\beta_1$ & $\beta_{n/2}$ & $\beta_{n}$ \\
		\hline
		$\theta=0.5$  & $ 94.5 $&$ 93.78 $&$ 95.24 $&$ 95.06 $&$ 95.42 $&$ 94.9 $&$ 95.12 $&$ 94.96 $  \\
		\hline
		$\theta=0$     & $94.96 $&$ 94.88 $&$ 95.14 $&$ 95 $&$ 95.16 $&$ 95.34 $&$ 94.82 $&$ 94.96 $
		\\
		\hline
		$\theta= -0.5$  &  $ 94.74 $&$ 95.42 $&$ 95.06 $&$ 95.28 $&$ 94.78 $&$ 95.24 $&$ 95.56 $&$ 94.46 $ \\
		\hline
	\end{tabular}
\end{table}

\subsection{Real data analysis}
In this section, we use the triple-dyad ratio method to analyze the Sina Weibo data collected by \cite{CAI2018network}. This dataset contains $4077$ individuals in an official MBA program, where a directed edge represents who follows whom. Because the explicit expressions of the triple-dyad ratio estimators depend on a logarithm ratio, we remove nodes that have a large influence on the estimators. For instance, when calculating $\hat{\theta}$, we remove such nodes the in-degree or out-degree of which is less than $5$, or values for which the $\mu_t^{abc}$ is zero. That is, we compute

\begin{align}\label{theta-hat}
	\frac{1}{m}\sum\limits_{t\in \Gamma } \log\frac{\sum\limits_{i,j\neq t}I_{it}^{01}I_{ij}^{00}I_{tj}^{01}}{\sum\limits_{i,j\neq t}I_{it}^{00}I_{ij}^{01}I_{tj}^{00}},
\end{align}
as the estimator of $\theta$, where $\Gamma:=\{t:\mu_t^{(abc)}>0,a,b,c=0,1, d_t \ge 5, b_t \ge 5\}$ and $|\Gamma|=m$. The same set $\Gamma$ is also applied to compute other estimators. The set $\Gamma$ in this dataset contains $560$ nodes.

\begin{figure}[!htp]
	\centering
	\subfigure[$\hat{\alpha}$]{
		\includegraphics[height=2.4in,width=2.55in, angle=0]{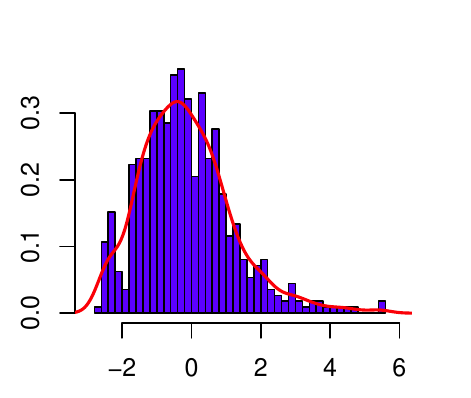}}
	\subfigure[$\hat{\beta}$]{
		\includegraphics[height=2.4in,width=2.8in, angle=0]{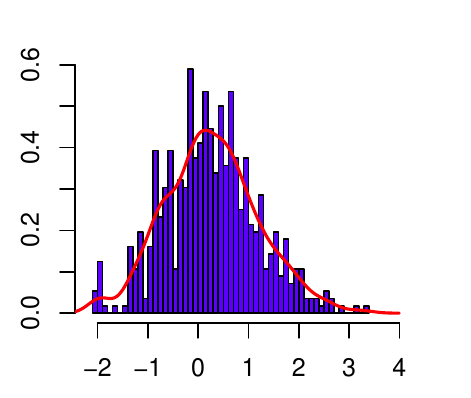}}
	\caption{Histogram of $\hat{\alpha}$ and $\hat{\beta}$. The red color indicates the density estimator.}\label{fig-alpha-real}
\end{figure}

We obtain that $\hat{\theta}$ and $\hat{\rho}$ are $-6.06$ and $7.56$, respectively. The value of $\hat{\theta}$ indicates that the Sina Weibo data are a sparse network. The minimum, median, and maximum values for $\hat{\alpha}_i$ are $-2.64$, $-0.27$, and $5.44$, while those for $\hat{\beta}$ are $-2.08$, $0.23$, and $3.32$, respectively. Thus, the network has a strong degree of heterogeneity. The histograms of $\hat{\alpha}$ and $\hat{\beta}$ are illustrated in Figures \ref{fig-alpha-real}. Computing the p-value to test for the existence of any reciprocity effects under $H_0:\rho=0$ yields a $p$-value of $0.032$, confirming a significant reciprocity effect.

\section{Discussion}
\label{section-discussion}
In this study, we proposed a triple-dyad ratio method for estimating $2n+2$ unknown parameters in the $p_1$ model. The estimator has an explicit expression, and thus is very easy to compute, unlike the MLE that requires iterative algorithms.
We also established consistency and asymptotic normality of the triple-dyad ratio estimator in response to the limitation that the asymptotic properties of the MLE are yet unknown. Our asymptotic theories can be used to  construct approximate confidence intervals for unknown parameters and to obtain approximate p-values for problems relating to hypothesis testing, including testing for a reciprocity effect or equality of degree parameters.

Our conditions imposed on the parameters to guarantee asymptotic theories may not be the best.
Our simulation studies show that the triple-dyad ratio estimator still has good asymptotic normal approximation when the network density is small, in the order of $n^{-1/3}$.
This indicates that the conditions might be relaxed. Nevertheless, the asymptotic behaviors of the estimators do not only depend on the range of parameters but also
the configuration of all the parameters. It would be of interest to see whether the condition could be improved.

We note that the expression in \eqref{thetaabk}, in terms of the logarithm of the ratio of probabilities of observing
two different subgraphs with exactly three nodes, is not unique.
That is, there exist other pairs of different subgraphs that could yield $\theta+\alpha_t+\beta_t$ as in \eqref{thetaabk}.
There are over $10$ non-isomorphic subgraphs with three nodes in directed networks. 
For dense networks, the accuracy of estimation among different kinds of ratios are similar, where there is no one that is optimal.
For sparse networks, the counts of the observed subgraphs with relatively more edges are much less than those with relatively less edges.
In view of this, we select the subgraphs used in \eqref{thetaabk} with only one or two edges in each subgraph.
For subgraphs with $4$ or more nodes, it is also possible to find such pairs such that $\theta+\alpha_t+\beta_t$ can be represented as the logarithm of the ratio of two probabilities observing different
subgraphs. However, the analysis will become much more tedious since there will be much more terms to analyze and also more complex dependent relationships.
\cite{Feng2026optimal} investigate the optimal estimator for the reciprocity parameter in sparse networks.
It is of interest to investigate whether there are optimal methods for estimating 
all parameters in the $p_1$ model in both dense and sparse networks.
We would like to investigate this issue in the future.

\bibliography{ref}
\bibliographystyle{apalike}

\end{document}